\documentclass[twocolumn,trackchanges]{aastex63}
\usepackage{amsmath}

\received{June 26, 2020}
\revised{June 30, 2020}
\accepted{\today}
\submitjournal{AJ}
\shorttitle{calibrating period-luminosity relation from the maximum-likelihood technique}
\shortauthors{Lazovik et al.}
\graphicspath{{./}{figures/}}
\begin{document}
\title{Calibrating the Galactic Cepheid Period-Luminosity relation\\ from the maximum-likelihood technique}

\correspondingauthor{Yaroslav Lazovik}
\email{yaroslav.lazovik@gmail.com}

\author{Yaroslav A. Lazovik}
\affiliation{Lomonosov Moscow State University, Faculty of Physics, 1 Leninskie
   Gory, bldg.2, Moscow, 119991, Russia}
\affiliation{Lomonosov Moscow State University, Sternberg Astronomical
Institute, 13 Universitetskiy prospect, Moscow, 119234, Russia}

\author{Alexey S. Rastorguev}
\affiliation{Lomonosov Moscow State University, Faculty of Physics, 1 Leninskie
   Gory, bldg.2, Moscow, 119991, Russia}
\affiliation{Lomonosov Moscow State University, Sternberg Astronomical
Institute, 13 Universitetskiy prospect, Moscow, 119234, Russia}

\begin{abstract}

We present a realization of the maximum-likelihood (ML) technique, which is one of the latest modifications of the Baade--Becker--Wesselink (BBW) method. Our approach is based on non-linear calibrations of the effective temperature and bolometric correction which take into account metallicity and surface gravity. It allows one to estimate the key Cepheid parameters, the distance modulus, and the interstellar reddening, combining photometric and spectroscopic data (including the effective temperature data). This method is applied to a sample of 44 Galactic Cepheids, for which multiphase temperature measurements are available. The additional data correction is performed to subtract the impact of the component in binary/multiple systems. We also study the effect of shock waves, whose presence in the stellar atmosphere distorts the observational data and leads to systematic errors in the obtained parameters. We determine the optimal restriction on the input data to eliminate this effect. This restriction provides accurate period-radius and period-luminosity relations which are consistent with the results in previous studies. We found the following relations: $log\,R = (0.68 \pm 0.03) \cdot log\,P + (1.14 \pm 0.03)$, $M_v = - (2.67 \pm 0.16) \cdot (log\,P - 1) - (4.14 \pm 0.05)$.

\end{abstract}


\section{Introduction} \label{sec:intro}
In the modern astronomy Cepheid variables play a particularly important role. Since the discovery of the period-luminosity  relation (PL, or also Leavitt law, \citealt{Leavitt+1908}; \citealt{Leavitt+1912}) in 1912, these stars have become the key objects in the context of extragalactic distance scale calibration and the Hubble constant estimation (\citealt{Riess+2011,Riess+2018, Riess+2019}). Decades of work led to a great progress in this field of research owing to both theory and observations. Technology development provided more accurate observational data while fundamental research in astrophysics prepared comprehensive theoretical background. However, calibrating precise PL relation still remains one of the priority astronomical goals.

Nowadays there are several methods used to solve this task and each of them has its own features and limitations. One of the most commonly used methods is the method of trigonometric parallax, which is inextricably linked to the Gaia mission (\citealt{Gaia}). However, the derived PL relation strongly depends on the parallax zero-point offset (\citealt{Groenewegen+2018}). Besides, in the case of Gaia DR2 data, astrometric precision achieved for close systems is low because such systems are not resolved, regardless of secondary brightness (\citealt{Ziegler+2018}). The capabilities of trigonometric parallax are very sensitive to the characteristic distance values, as the astrometric precision delivered for far located stars is much lower. The distances obtained for Cepheids in open clusters are more reliable, but sometimes it is not straightforward to confirm cluster membership. Eventually, the limited number of such objects prevents calibrating precise PL relation based on cluster Cepheids only.

In light of the above, the Baade--Becker--Wesselink (BBW; \citealt{Baade+1926}; \citealt{Becker+1940}; \citealt{Wesselink+1946}) method stands in the foreground as it is devoid of the mentioned flaws. Universality of this method makes it the only reasonable way to establish extragalactic distance scale with Cepheid variables. Nowadays the BBW method is more complicated than it was in the original works, many different modifications have been proposed, among which the infrared surface-brightness (IRSB;  \citealt{Barnes+Evans+1976}) technique deserves special attention as the most frequently used implementation. Nevertheless, it is not the only approach. In this study we present another modification of the BBW method, namely the maximum-likelihood (ML) technique, whose basics were firstly described by \cite{Balona+1977}, that's why we also call it the Baade--Becker--Wesselink--Balona (BBWB) method. The generalization of this method has been developed by Rastorguev and Dambis (RD version; \citealt{Rastorguev+Dambis+2010}). The key point of our approach is using the multiphase effective temperature data in order to independently determine the stellar distance and the main physical parameters, such as radius and luminosity, as well as the amount of interstellar reddening. IRSB and ML techniques have many common features since they both rely on the identical theoretical framework, but consider the same task from different angles. It will be demonstrated that the ML technique has advantages over the IRSB method. Today the capabilities of our approach are limited by the amount of observational material, in the present study we work with relatively small sample consisting of 44 Galactic Cepheids, for which the multiphase effective temperature variations are available. However, it will be shown that the ML technique has a potential to become a useful tool in the context of distance scale calibration, as it is physically based and independent of other geometric techniques used to investigate the intrinsic properties of Cepheids. By intercomparing these results we are certain to learn more about Cepheids as physical systems locally and learn more about the physical expansion of the universe by the application of Cepheids (and their PL relation) to the more distant universe.

This paper is structured as follows. In the next section we emphasize the theoretical basis of the ML method. In Section~\ref{sec:data} we dwell on the observational data and data reduction. The results are presented in Section~\ref{sec:results}, the discussion is given in Section~\ref{sec:discussion}. Finally, we summarize our work in Section~\ref{sec:summary}.
%\vfill\null - % to break the column

\section{Method} \label{sec:method}
We now briefly outline the RD version of the ML technique (see \cite{Rastorguev+Dambis+2010}, \cite{Rastorguev+2013} and \cite{Rastorguev+Lazovik} for details). The central equation of this method is the relation for the model light curve which can be derived from the Stefan--Boltzmann law and the relation between absolute magnitude and apparent magnitude:
\begin{equation}\label{eq1}
\ m = Y - 5 \cdot \log{\frac{R}{R_\odot}} + \Psi,
\end{equation}
where $Y$ is a constant depending on stellar apparent distance:
\begin{equation}\label{eq2}
\ Y = (m - M)_{app} + M_{bol\odot} +  10 \cdot\log{T_{eff\odot}},
\end{equation}
and $\Psi$ is a function of normal color index $CI_0=CI-CE$ ($CE$ is
the color excess):
\begin{equation}\label{eq3}
  \ \Psi(CI_0) = BC +  10 \cdot \log{T_{eff}},
\end{equation}
where $BC$ is the bolometric correction. The value of $\Psi$ can be expressed from non-linear calibrations $BC(CI_0)$ and $\log{T_{eff}}(CI_0)$:
\begin{equation}\label{eq4}
  \ BC = a_0 + \sum_{k=1}^{N_1} a_k CI_0^{k}.
\end{equation}
\begin{equation}\label{eq5}
  \ \log{T_{eff}} = b_0 + \sum_{k=1}^{N_2} b_k CI_0^{k}.
\end{equation}
Note that we also take into account the impact of surface gravity and metallicity, as the coefficients  $b_k$ in Eq.~\ref{eq5} depend on $log\,g$ and $[Fe/H]$. Both $log\,g$ and $[Fe/H]$ are assumed as constant parameters.

\begin{figure}[ht!]
\plotone{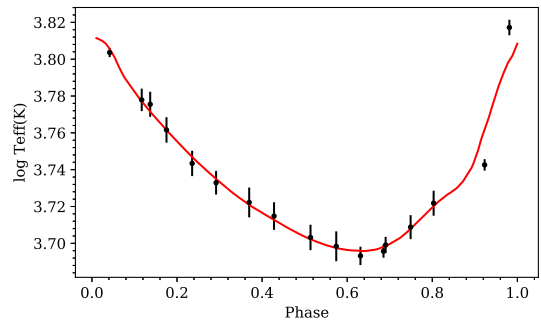}
\caption{Effective temperature curve for CD Cyg. Dots with error bars: effective temperatures from \cite{Luck+2018}. Red solid line: the model curve calculated using the relation from the present work (see \cite{Rastorguev+Lazovik})    \label{fig:fig1}}
\end{figure}

Requiring the best agreement between the observed values of the effective temperature and the model values computed from Eq.~\ref{eq5}, we estimate the color excess ($E(B-V)$), which is then used to derive $\Psi$. We initially employed $BC((B - V)_0)$ relation from \cite{Flower} and $\log{T_{eff}}((B - V)_0)$ relation from \cite{Bessell+1998}. After estimating $E(B-V)$ values for 33 Cepheids we re-calibrated $\log{T_{eff}}((B - V)_0)$ relation, treating Bessell's coefficients as the first approximation. The final $\log{T_{eff}}((B - V)_0)$ expression is given by \cite{Rastorguev+Lazovik}. An example of the effective temperature model curve for CD Cyg is presented in Figure~\ref{fig:fig1}.

We obtained the radius variation $\Delta R(\varphi)$, integrating radial-velocity curve over time:
\begin{equation}\label{eq6}
  \ \Delta R(\varphi) = - p \cdot \int_{\varphi_0}^{\varphi}(V_r(\varphi) - V_{\gamma} ) \frac{P}{2\pi} d\varphi ,
\end{equation}
where $p$ is the projection factor; $V_r(\varphi)$ is the radial velocity; $V_{\gamma}$ is the systematic radial velocity; $R_0$ is the average radius value; $P$ is the pulsation period; and $\varphi$ is
the current phase of the radial velocity curve. The main uncertainty of our method arises from the projection factor (p-factor) estimation. P-factor provides a conversion from radial to pulsation velocity. \cite{Nardetto+2017} decomposes p-factor into three components: geometric projection factor ($p_0$), the atmospheric velocity gradient ($f_{grad}$), and the relative motion of
the optical pulsating photosphere with respect to the corresponding mass elements ($f_{o-g}$).  The authors propose different values of p-factor,  there's still no consensus concerning its correlation with the pulsation period (\citealt{Nardetto+2004, Nardetto+2007, Nardetto+2009, Groenewegen+2007}). Moreover, for a given Cepheid, the projection factor may change during the pulsation cycle (\citealt{Hindsley,Gautschy,Butler,Sasselov, Sabbey}). Today we lack quantitative theoretical estimates for such variations, that's why we neglect them. In the present work we adopt relation from \cite{Nardetto+2007}:
\begin{equation}\label{eq7}
  \ p = 1.376 - 0.064 \cdot log\,P
\end{equation}
The mean radius value is derived from the main Balona equation:
\begin{equation}\label{eq8}
  \ m = c_0 +  \sum_{k=1}^{N_3} c_k CI^{k}  - 5 \cdot \log(R_0 + \Delta R(\varphi)),
\end{equation}
  where $c_k$ and $R_0$ are the unknown parameters.
  
\begin{figure}[ht!]
\plotone{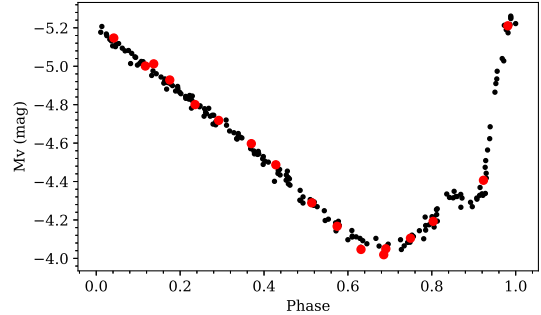}
\caption{Light curves for CD Cyg. Black dots: the observed light curve shifted by the value of the apparent distance modulus. Red circles: values calculated from Eq.(\ref{eq1})      \label{fig:fig2}}
\end{figure}

Once we obtained the stellar color excess and radius, the apparent distance remains the only unknown parameter in the light curve equation (Eq.~\ref{eq1}), which can be easily found using the least-squares method. The last step of our algorithm is calculating the absolute distance modulus:
\begin{equation}\label{eq9}
  \ (m - M)_0 = (m - M)_{app} - A,
\end{equation}
where interstellar extinction $A$ can be determined as $A_{\lambda} = R_{\lambda} \cdot E(B-V)$, where $R_{\lambda}$ is the total-to-selective extinction ratio for the passband–color pair considered ($R_v = 3.3$; \citealt{Storm+2004}). Figure~\ref{fig:fig2} shows the observed and model V-band light curves for CD Cyg.

\begin{figure}[ht!]
\plotone{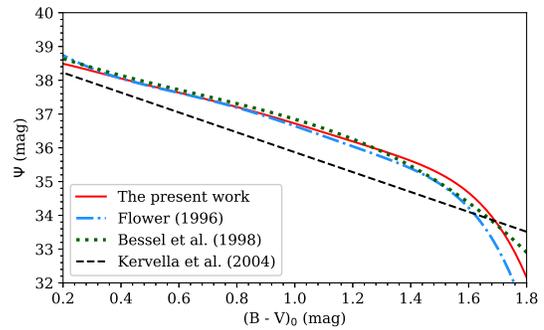}
\caption{IRSB and ML $\Psi(CI_0)$ calibrations for CD Cyg ($log\,g = 1.30 $; $[Fe/H] = 0.15$). Red solid line: relation used in the present work. Dashed-dotted blue line: relation from \cite{Flower}. Dotted green line: relation from \cite{Bessell+1998}.
Dashed black line: relation from \cite{Kervella} \label{fig:fig3}}
\end{figure}

As noted by \cite{Rastorguev+Lazovik}, the outlined method and the IRSB technique are based on the same theoretical material. However, if the IRSB technique may be considered as the simulation of radius changes, the ML technique simulates the light curve.

Contrary to IRSB technique, our method doesn't require preliminary color excess estimations. As explained by \cite{Madore}, the errors in the reddening determinations are the factor that increases the dispersion of the existing PL relations. Moreover, we found that the effective temperature calibrations are very sensitive to the color excess variations, that's why it's important to be able to directly estimate the reddenings for individual objects to achieve high precision of the distance scale. Adopting gravity- and metallicity-dependent non-linear calibrations leads to significantly different results with respect to the linear calibrations of the IRSB method. A comparison of the ML (\citealt{Flower,Bessell+1998}; the present work) versions with the IRSB version (\citealt{Kervella}) of $\Psi(CI_0)$ calibration in the case of CD Cyg is depicted in Figure~\ref{fig:fig3}.

In addition, the results obtained with the IRSB method differ from the results of the present research simply because of our choice in favor of B-band photometry over traditionally used infrared photometry. Such choice is justified by the fact that stellar radius, pulsation velocity, limb-darkening law, and projection factor are all wavelength-dependent (\citealt{Marengo, Nardetto+2009,Howarth,Neilson}). Given that the radial velocity measurements are related to the optical spectrum, solving the Balona equation (Eq.~\ref{eq8}) with infrared photometry (corresponding to the different radius value), affected by the radiation of circumstellar gas, without any transformations would introduce errors into the final solution. For this reason, in this study, we settle on using data which corresponds to the single part of spectrum and thus to the single radius value.

\startlongtable
\begin{deluxetable*}{cccccccc}
\tablenum{1}
\tablecaption{Parameters of 44 Cepheid variables obtained with [0.00; 0.85] phase constraint\label{tab1}}
\tablewidth{0pt}
\tablehead{
\colhead{} & \colhead{} & \colhead{} & \colhead{Fundamental period} & \colhead{$E(B-V)$} & \colhead{$<R>$} & \colhead{$\overline{M_v}$} & \colhead{$(m - M)_0$} \\
\colhead{Cepheid} & \colhead{Sample} & \colhead{Binary} & \colhead{($days$)} & \colhead{($mag$)} &  \colhead{($R_\odot$)} &  \colhead{($mag$)} &  \colhead{($mag$)}\\
}
\startdata
AW Per & 3 & Yes & 6.463 & 0.59 $\pm$ 0.01 & 39.2 $\pm$ 2.1 & -3.34 $\pm$ 0.12 & 8.96 $\pm$ 0.17\\
BB Her & 1 & No & 7.508 & 0.41 $\pm$ 0.02 & 56.2 $\pm$ 3.0 & -3.85 $\pm$ 0.06 & 12.60 $\pm$ 0.06 \\
BG Lac & 2 & No & 5.332 & 0.29 $\pm$ 0.01 & 41.8 $\pm$ 2.1 & -3.24 $\pm$ 0.07 & 11.18 $\pm$ 0.09 \\
CD Cyg & 1 & No & 17.074 & 0.59 $\pm$ 0.01 & 99.2 $\pm$ 2.0 & -4.85 $\pm$ 0.04 & 11.87 $\pm$ 0.13 \\
CF Cas & 1 & No & 4.875 & 0.54 $\pm$ 0.02 & 44.3 $\pm$ 1.2 & -3.34 $\pm$ 0.04 & 12.71 $\pm$ 0.12 \\
CV Mon & 3 & No & 5.379 & 0.69 $\pm$ 0.02 & 51.1 $\pm$ 2.7 & -3.77 $\pm$ 0.12 & 11.79 $\pm$ 0.12 \\
Delta Cep & 1 & Yes & 5.366 & 0.09 $\pm$ 0.02 & 43.7 $\pm$ 2.3 & -3.51 $\pm$ 0.06 & 7.17 $\pm$ 0.06 \\
DL Cas & 3 & Yes & 11.268 & 0.65 $\pm$ 0.02 & 87.0 $\pm$ 4.8 & -4.82 $\pm$ 0.12 & 11.70 $\pm$ 0.18 \\
DT Cyg & 3 & No & 3.520 & 0.04 $\pm$ 0.01 & 42.3 $\pm$ 4.7 & -3.63 $\pm$ 0.15 & 9.28 $\pm$ 0.15 \\
Eta Aql & 1 & No & 7.177 & 0.16 $\pm$ 0.01 & 56.3 $\pm$ 2.5 & -3.92 $\pm$ 0.06 & 7.29 $\pm$ 0.07 \\
FF Aql & 3 & Yes & 6.297 & 0.27 $\pm$ 0.01 & 53.7 $\pm$ 7.6 & -4.16 $\pm$ 0.18 & 8.64 $\pm$ 0.17 \\
FM Aql & 1 & No & 6.114 & 0.69 $\pm$ 0.02 & 52.2 $\pm$ 1.7 & -3.79 $\pm$ 0.05 & 9.80 $\pm$ 0.15 \\
FN Aql & 1 & No & 9.482 & 0.48 $\pm$ 0.02 & 62.3 $\pm$ 1.4 & -3.87 $\pm$ 0.04 & 10.67 $\pm$ 0.11 \\
RS Ori & 3 & No & 10.658 & 0.37 $\pm$ 0.01 & 72.7 $\pm$ 4.2 & -4.60 $\pm$ 0.12 & 11.78 $\pm$ 0.14 \\
RT Aur & 1 & No & 3.728 & 0.06 $\pm$ 0.01 & 36.3 $\pm$ 1.6 & -3.19 $\pm$ 0.06 & 8.44 $\pm$ 0.06 \\
RX Aur & 1 & No & 11.624 & 0.34 $\pm$ 0.01 & 72.5 $\pm$ 2.4 & -4.50 $\pm$ 0.05 & 11.03 $\pm$ 0.09 \\
RX Cam & 2 & Yes & 7.912 & 0.55 $\pm$ 0.01 & 46.8 $\pm$  2.7 & -3.52 $\pm$ 0.08 & 9.39 $\pm$ 0.14 \\
S Sge & 1 & Yes & 8.382 & 0.17 $\pm$ 0.01 & 50.6 $\pm$ 1.2 & -3.68 $\pm$ 0.04 & 8.74 $\pm$ 0.05 \\
S Vul & 1 & No & 68.438 & 1.15 $\pm$ 0.03 & 246.0 $\pm$ 8.1 & -6.89 $\pm$ 0.08 & 12.06 $\pm$ 0.25 \\
SS Sct & 2 & No & 3.671 & 0.38 $\pm$ 0.03 & 36.3 $\pm$ 0.9 & -3.13 $\pm$ 0.08 & 10.08 $\pm$ 0.11 \\
SU Cyg & 3 & Yes & 5.417 & 0.10 $\pm$ 0.02 & 49.8 $\pm$ 5.8 & -3.98 $\pm$ 0.16 & 10.55 $\pm$ 0.16 \\
SV Mon & 1 & No & 15.235 & 0.29 $\pm$ 0.02 & 93.0 $\pm$ 1.7 & -4.63 $\pm$ 0.06 & 11.93 $\pm$ 0.08 \\
SV Vul & 1 & No & 44.969 & 0.62 $\pm$ 0.03 & 192.0 $\pm$ 3.3 & -6.08 $\pm$ 0.06 & 11.24 $\pm$ 0.14 \\
T Mon & 1 & Yes & 27.033 & 0.30 $\pm$ 0.05 & 119.7 $\pm$ 1.8 & -4.95 $\pm$ 0.05 & 10.09 $\pm$ 0.08 \\
T Vul & 1 & Yes & 4.435 & 0.07 $\pm$ 0.02 & 39.8 $\pm$ 1.5 & -3.33 $\pm$ 0.05 & 8.88 $\pm$ 0.05 \\
TT Aql & 1 & No & 13.755 & 0.59 $\pm$ 0.02 & 87.0 $\pm$ 1.9 & -4.58 $\pm$ 0.06 & 9.77 $\pm$ 0.14 \\
U Aql & 3 & Yes & 7.024 & 0.44 $\pm$ 0.02 & 41.4 $\pm$ 1.7 & -3.36 $\pm$ 0.14 & 8.33 $\pm$ 0.17 \\
U Sgr & 1 & No & 6.745 & 0.46 $\pm$ 0.01 & 48.9 $\pm$ 2.2 & -3.62 $\pm$ 0.06 & 8.81 $\pm$ 0.11 \\
U Vul & 2 & Yes & 7.990 & 0.72 $\pm$ 0.02 & 40.9 $\pm$ 1.2 & -3.35 $\pm$ 0.05 & 8.10 $\pm$ 0.15 \\
V500 Sco & 2 & No & 9.317 & 0.62 $\pm$ 0.03 & 62.1 $\pm$ 3.9 & -4.05 $\pm$ 0.11 & 10.75 $\pm$ 0.17 \\
VX Per & 2 & No & 10.885 & 0.53 $\pm$ 0.02 & 79.7 $\pm$ 3.2 & -4.60 $\pm$ 0.07 & 12.14 $\pm$ 0.13 \\
W Gem & 2 & No & 7.914 & 0.30 $\pm$ 0.03 & 47.1 $\pm$ 1.6 & -3.60 $\pm$ 0.08 & 9.56 $\pm$ 0.10 \\
W Sgr & 1 & Yes & 7.595 & 0.13 $\pm$ 0.01 & 48.8 $\pm$ 1.5 & -3.66 $\pm$ 0.06 & 7.92 $\pm$ 0.06 \\
WZ Sgr & 1 & No & 21.850 & 0.59 $\pm$ 0.03 & 120.2 $\pm$ 2.1 & -5.03 $\pm$ 0.06 & 11.12 $\pm$ 0.14 \\
X Cyg & 1 & No & 16.386 & 0.35 $\pm$ 0.02 & 95.9 $\pm$ 2.5 & -4.59 $\pm$ 0.07 & 9.84 $\pm$ 0.10 \\
X Pup & 3 & No & 25.965 & 0.53 $\pm$ 0.03 & 106.8 $\pm$ 2.8 & -5.14 $\pm$ 0.12 & 11.93 $\pm$ 0.16 \\
X Vul & 1 & No & 6.320 & 0.85 $\pm$ 0.02 & 49.4 $\pm$ 2.0 & -3.71 $\pm$ 0.05 & 9.73 $\pm$ 0.18 \\
XX Sgr & 2 & No & 6.424 & 0.58 $\pm$ 0.02 & 55.3 $\pm$ 2.7 & -4.07 $\pm$ 0.07 & 11.02 $\pm$ 0.14 \\
Y Lac & 3 & No & 6.090 & 0.15 $\pm$ 0.02 & 48.7 $\pm$ 2.3 & -3.77 $\pm$ 0.11 & 12.42 $\pm$ 0.12 \\
Y Oph & 2 & No & 17.128 & 0.78 $\pm$ 0.01 & 103.5 $\pm$ 3.0 & -5.29 $\pm$ 0.06 & 8.87 $\pm$ 0.17 \\
Y Sgr & 2 & No & 5.773 & 0.23 $\pm$ 0.02 & 48.7 $\pm$ 1.1 & -3.66 $\pm$ 0.04 & 8.62 $\pm$ 0.06 \\
YZ Sgr & 1 & No & 9.554 & 0.36 $\pm$ 0.01 & 55.7 $\pm$ 1.5 & -3.81 $\pm$ 0.04 & 9.97 $\pm$ 0.08 \\
Z Lac & 1 & Yes & 10.886 & 0.49 $\pm$ 0.02 & 70.1 $\pm$ 1.1 & -4.29 $\pm$ 0.06 & 11.10 $\pm$ 0.12 \\
Zet Gem & 3 & No & 10.150 & 0.07 $\pm$ 0.01 & 67.8 $\pm$ 3.3 & -4.08 $\pm$ 0.12 & 7.71 $\pm$ 0.12 \\
\enddata
\tablecomments{choice of [0.00; 0.85] phase restriction is explained in Section~\ref{sec:shock waves} }
\end{deluxetable*}

\section{The data} \label{sec:data}
In this study we use extensive multicolor photoelectric and CCD photometry in B and V bands from \cite{Berdnikov}. As for spectroscopic data, the radial velocity measurements come from \cite{Gorynya+1992,Gorynya+1996,Gorynya+1998,Gorynya+2002}, the effective temperature and metallicity measurements are from \cite{Luck+2018}.
The data are cleaned as we identify and remove the data sets with the highest dispersion relative to the phase curve. We also exclude the observations corresponding to the earliest epochs to ensure that the observational data are synchronous in time to prevent any systematic errors coming from evolutionary period changes resulting in phase shifts between light, color, and radial-velocity variations. The period has been subsequently recalculated with the remaining data.
For data interpolation we apply locally-estimated scatterplot smoothing (loess). We found that loess regression algorithm provides smoother fits than Fourier series, although this difference doesn't affect the final PR and PL relations.

\subsection{Cepheids in binary and multiple systems} \label{subsec:systems}
\begin{figure}[ht!]
\plotone{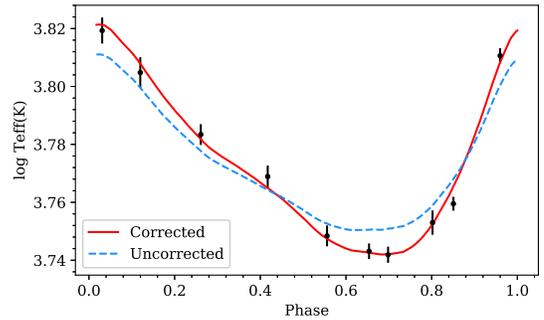}
\caption{Effective temperature curve for the binary Cepheid AW Per. Dots: effective temperatures from \cite{Luck+2018}. Blue dashed line: the model curve corresponding to the uncorrected photometric data. Red solid line: the model curve corresponding to the corrected photometric data   \label{fig:fig4}}
\end{figure}
The full sample of stars considered in our research consists of 44 Galactic Cepheids, among which there are objects in binary and multiple systems. For these Cepheids the additional algorithm has been carried out to correct data from the impact of the component. Firstly, we calculate the orbital parameters for the case of binary system assuming the minimal dispersion of residual velocities (corresponding to the pulsating motion) on the phase curve, after which we subtract orbital velocity from radial velocity data. Secondly, we subtract the impact of the brightest component from the photometric data. For this purpose, we take spectral type of the component from \cite{Evans+1992,Evans+1995,Evans+2013,Udalski}. With given spectral type, using stellar parameters from \cite{Pecaut+2013} and the apparent distance modulus calculated with the initial (uncorrected) photometry, we compute B and V apparent magnitudes of the component, which allows us to correct the Cepheid light curves. Applying photometric correction doesn't lead to significant changes in the obtained stellar parameters for most of the Cepheids, however, there are several exceptions. The example of AW Per, shown in Figure~\ref{fig:fig4}, demonstrates that the effective temperatures calculated with modified color index fit the observational data better. It is important to note that AW Per doesn't belong to the list of 33 Cepheids which were used to refine $\log{T_{eff}}((B-V)_0)$ calibration.

\subsection{Samples}

All Cepheids studied in this research are divided into three samples depending on the expected precision of the final parameters. Several factors have been taken into account, including the data quality and completeness, the presence of the component and its impact on the derived solution, the pulsation type (overtone or fundamental). The first sample, corresponding to the Cepheids with the most reliable solutions, contains 23 objects, the second sample contains 10 objects, and the third sample contains 11 objects. The list of Cepheids, sample membership, and the presence of the component are given in Table~\ref{tab1}.

\begin{deluxetable*}{cccccccc}
\tablenum{2}
\tablecaption{Period-radius relations and period-luminosity relations in the form $log(R) = a_r \cdot  log(P)  + b_r$ and $M_v = a_v \cdot  (log(P) - 1)  + b_v$, respectively \label{tab2}}
\tablewidth{0pt}
\tablehead{
\colhead{Phase restriction} & \colhead{Samples} & \colhead{$a_r$} & \colhead{$b_r$} & \colhead{$SD_r$} & \colhead{$a_v$} & \colhead{$b_v$} &  \colhead{$SD_v$} \\
}
\startdata
No restriction [0.00; 1.00] & Sample 1 & 0.67 $\pm$ 0.03  & 1.15 $\pm$ 0.03   & 0.03 & -2.51 $\pm$ 0.18  & -4.19 $\pm$ 0.05  & 0.24 \\
No restriction [0.00; 1.00] & Sample 1, 2 & 0.66 $\pm$ 0.03 & 1.17 $\pm$ 0.03 & 0.04 & -2.50 $\pm$ 0.18 & -4.22 $\pm$ 0.05 & 0.25\\
No restriction [0.00; 1.00] & Sample 1, 2, 3 & 0.65 $\pm$ 0.02 & 1.18 $\pm$ 0.03 & 0.04 & -2.48 $\pm$ 0.16 & -4.23 $\pm$ 0.04 & 0.27 \\
Restriction [0.00; 0.85] & Sample 1  &  0.68 $\pm$ 0.03 &  1.14 $\pm$ 0.03 & 0.03 & -2.65 $\pm$ 0.15 & -4.13 $\pm$ 0.05 & 0.20 \\
Restriction [0.00; 0.85] & Sample 1, 2  & 0.68 $\pm$ 0.03 &  1.14 $\pm$ 0.03 & 0.04 & -2.67 $\pm$ 0.16 & -4.14 $\pm$ 0.05 & 0.25\\
Restriction [0.00; 0.85] & Sample 1, 2, 3 & 0.68 $\pm$ 0.02 & 1.14 $\pm$ 0.03  & 0.05 & -2.65 $\pm$ 0.15 & -4.15 $\pm$ 0.04 & 0.30\\
\enddata
\tablecomments{The coefficients are derived using the weighted least squares method. $SD_r$ refers to standard deviation of the period-radius relation, $SD_v$ refers to standard deviation of the period-luminosity relation. We recommend relations computed for the combination of the first and the second samples}
\end{deluxetable*}

\section{Results} \label{sec:results}

\subsection{Overtone Cepheids} \label{sec:ov}

Obtaining the PR relation allows one to identify overtone pulsators, which are shifted towards the lower period values relative to the linear fit. The fundamental period of these Cepheids can be calculated as a multiplication of the observed period and the constant coefficient (we assign it a value of 1.41). We suspect six objects to be the first overtone Cepheids, namely DL Cas, DT Cyg, FF Aql, RS Ori, SU Cyg, Y Lac. Only two of them belong to the list of the first overtone pulsators in the General Catalogue of Variable Stars (\citealt{Samus}), namely DT Cyg and FF Aql. The first overtone pulsation mode of DT Cyg is also suspected by \cite{Ferro}, while \cite{Udalski} discuss the probability of FF Aql being an overtone pulsator. Besides the above two stars, RS Ori is considered to be overtone Cepheid in the Gaia DR2 catalogue (\citealt{Gaia}). Because we can't clarify whether the shift in the PR diagram is caused by the different type of pulsation, by the uncertainties of the observational data or by the possible evolutionary effects, these stars have been included in the third sample of objects. Moreover, the applicability of our calibrations to the overtone pulsators is questionable, that's why we suggest that the derived parameters of the stars listed above might contain additional errors.

Linear expressions of PR and PL relations are reported in Table~\ref{tab2}.

\subsection{Shock waves} \label{sec:shock waves}

During the end of the pulsation cycle, which is related to the rebound around the minimum radius, shock waves arise in the atmosphere of Cepheids. Using high-resolution optical and infrared spectra and synthetic line profiles, \cite{Sabbey} has demonstrated that the presence of shock waves introduces asymmetries in the Cepheid line profiles. These asymmetries result in significant systematic errors in radial velocity and effective temperature measurements. Besides, the calibrations adopted in this work might not be applicable to the phase region corresponding to the formation of shock waves. As a consequence, this phase region should be disregarded while calculating radius, color excess, and apparent distance modulus.

\begin{figure}[ht!]
\plotone{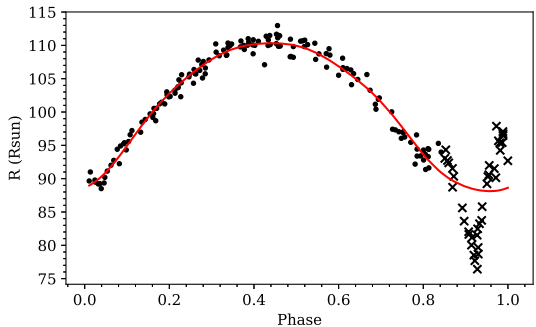}
\caption{Pulsation curve for CD Cyg. Dots and crosses: radius estimations from Eq.~\ref{eq8}. Dots belong to [0.00; 0.85] phase interval, crosses belong to [0.85; 1.00] phase interval (disregarded region). Red solid line: integrated radial velocity curve   \label{fig:fig5}}
\end{figure}

Figure~\ref{fig:fig5} illustrates mismatch between the pulsation curve obtained by integrating radial velocity curve and radius estimations from the Balona equation (Eq.~\ref{eq8}). The Balona equation, which is based on $\log{T_{eff}}(CI_0)$ relation, doesn't give satisfactory result during shock waves formation, indicating that this relation alone has its own application boundaries for Cepheid variables.

\begin{figure}[ht!]
\plotone{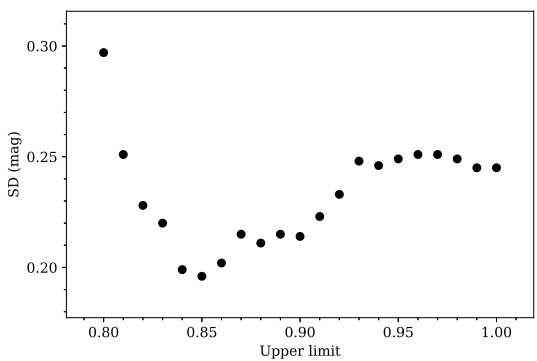}
\caption{Standard deviation of the first sample Cepheids in the PL diagram plotted against upper phase limit \label{fig:fig6}}
\end{figure}

\begin{figure*}[ht!]
\plottwo{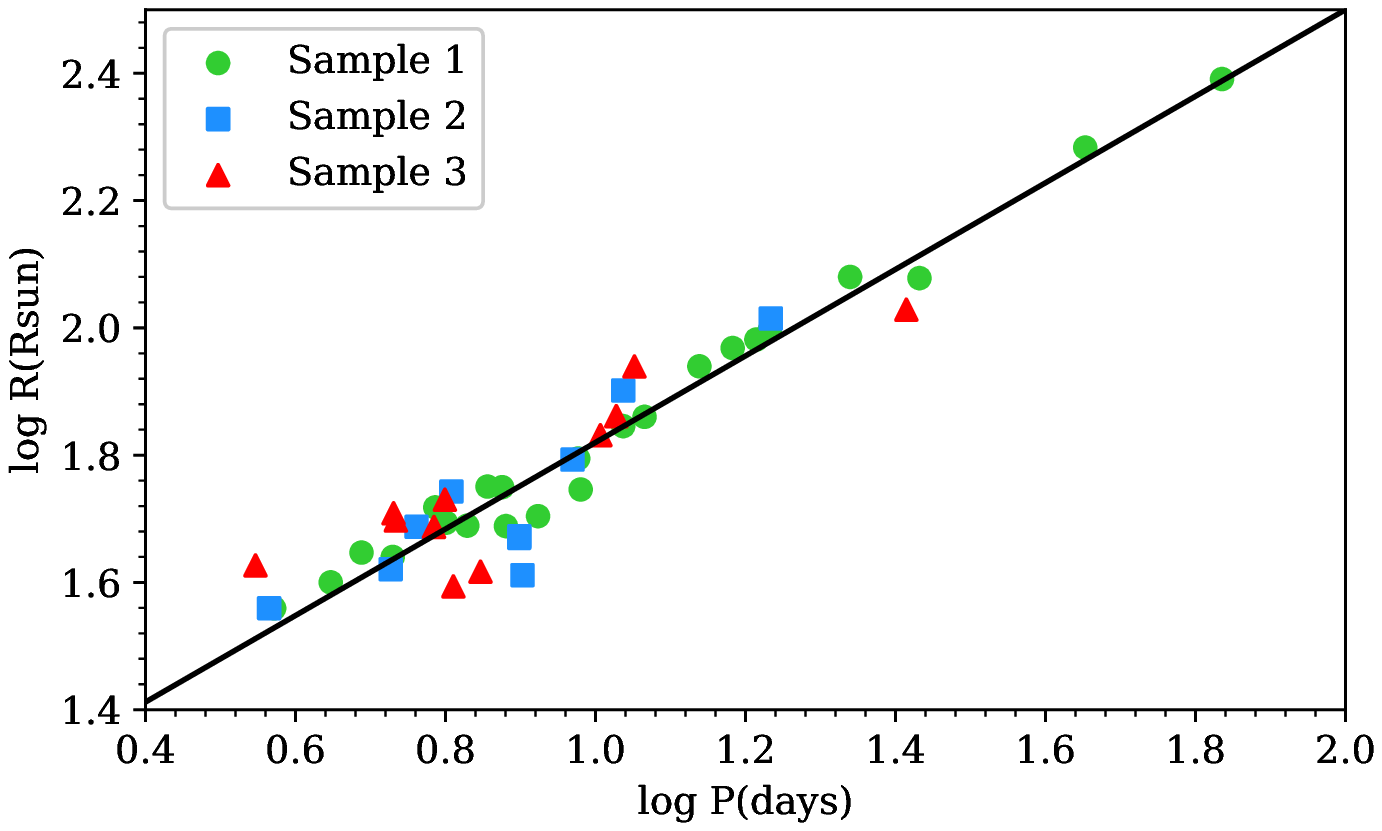}{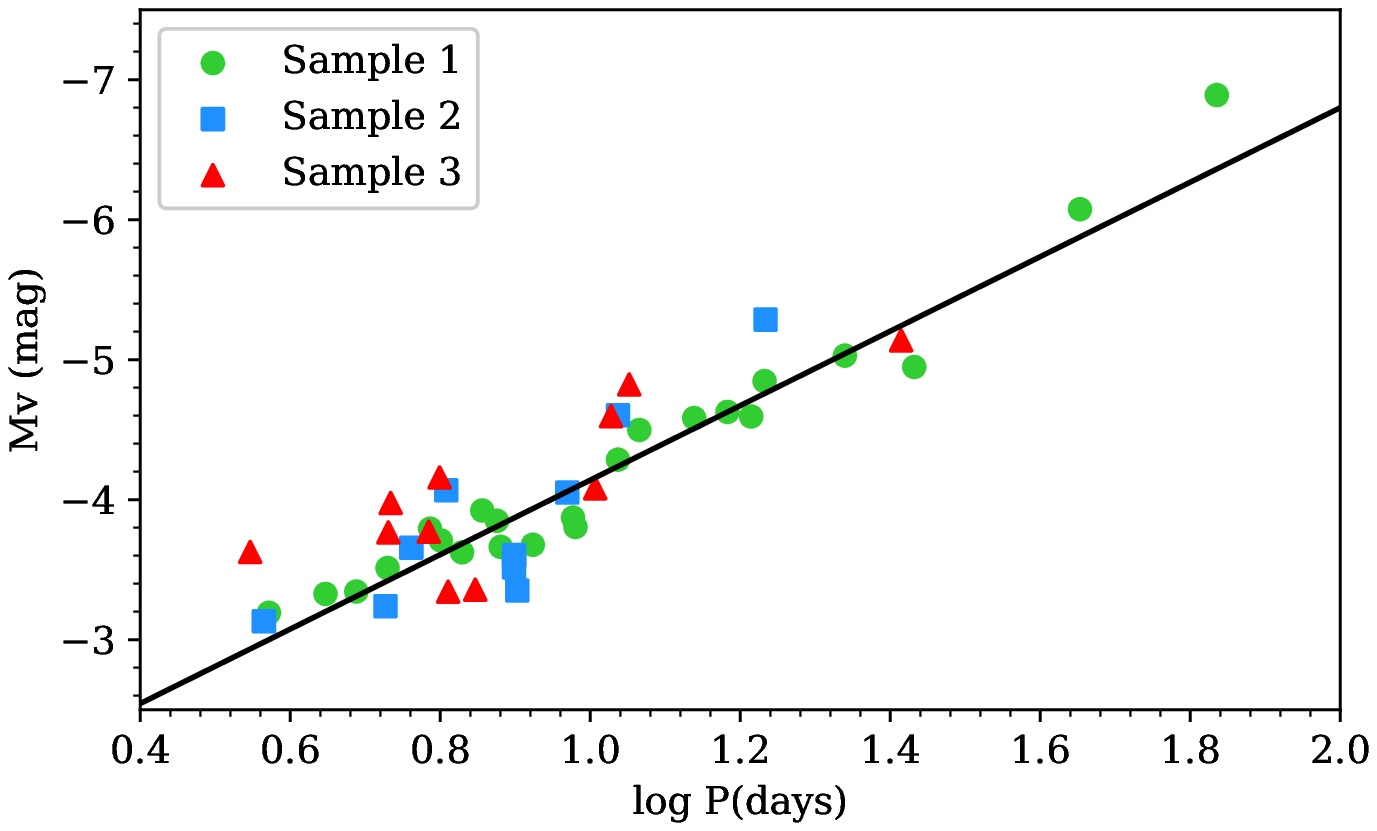}
\caption{\textbf{Left panel}: period-radius diagram. \textbf{Right panel}: period-luminosity diagram in the V band. Green circles are the first sample Cepheids, blue squares are the second sample Cepheids, red triangles are the third sample Cepheids. Solid line is the linear fit.     \label{fig:fig7}}
\end{figure*}

To obtain the application boundaries we consider the first sample of Cepheids and the upper limits in the range [0.80; 1.00] with a step of 0.01 to find out which constraint leads to the smallest scatter relative to linear PL relation. Phase restrictions are applied to all Cepheids, despite the fact that some of them don't exhibit the presence of shock waves. Some authors (for example, \citealt{Storm+2011}) disregard [0.8, 1.0] phase interval to avoid using distorted data. However, we showed in \cite{Lazovik} that phase constraint substantially affects the adopted PL relation, meaning that careful deduction of optimal restriction has to be provided. As shown in Figure~\ref{fig:fig6}, excluding data inside [0.85; 1.00] minimizes spread in the PL diagram. Linear expressions of the PR and PL relations are given in Table~\ref{tab2}, PR and PL diagrams are plotted in Figure~\ref{fig:fig7}. We recall that the parameters calculated for Cepheids of the third sample are less reliable, we therefore recommend using relations derived from a combination of the first and the second samples.

\subsection{Instability strip} \label{sec:gr}

\begin{figure*}[ht!]
\plottwo{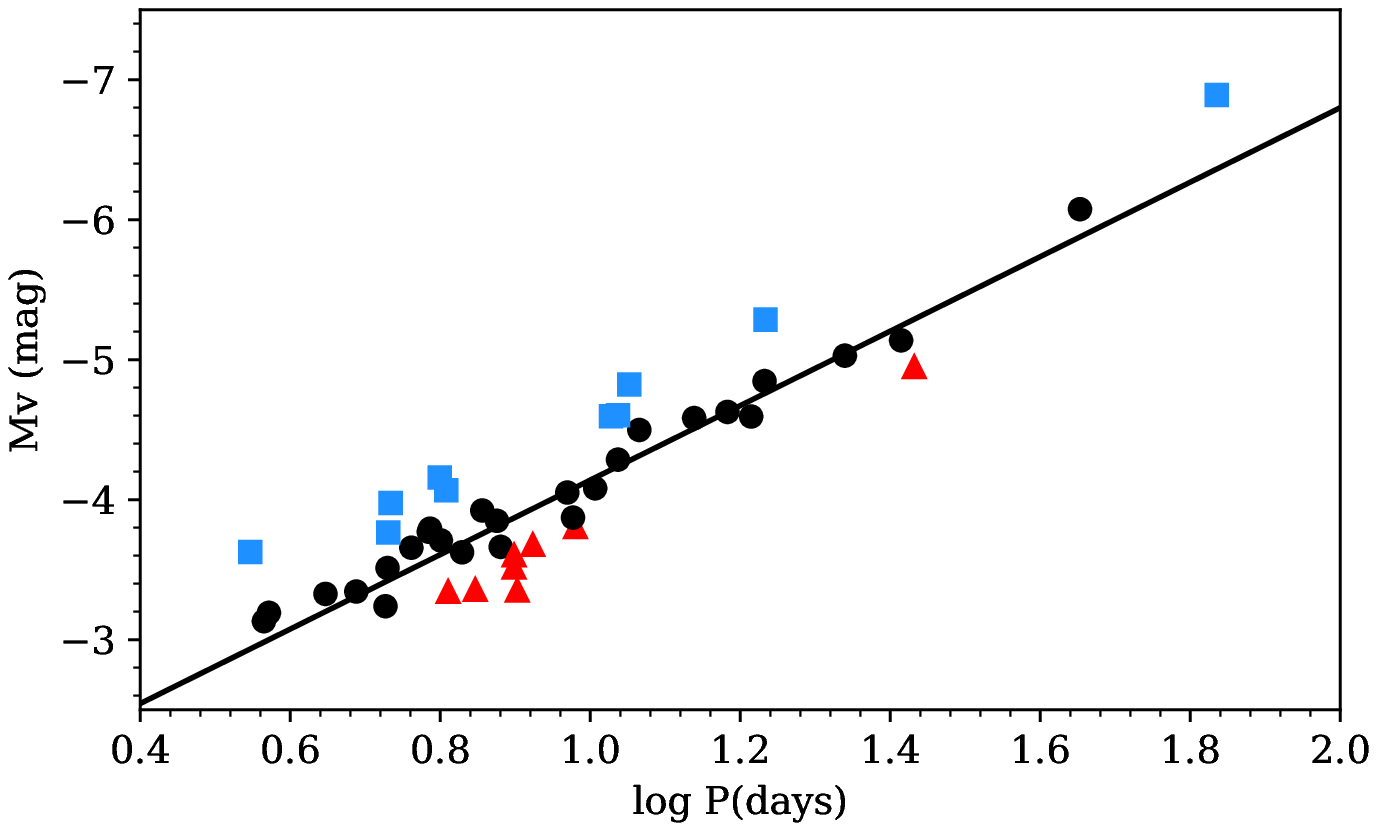}{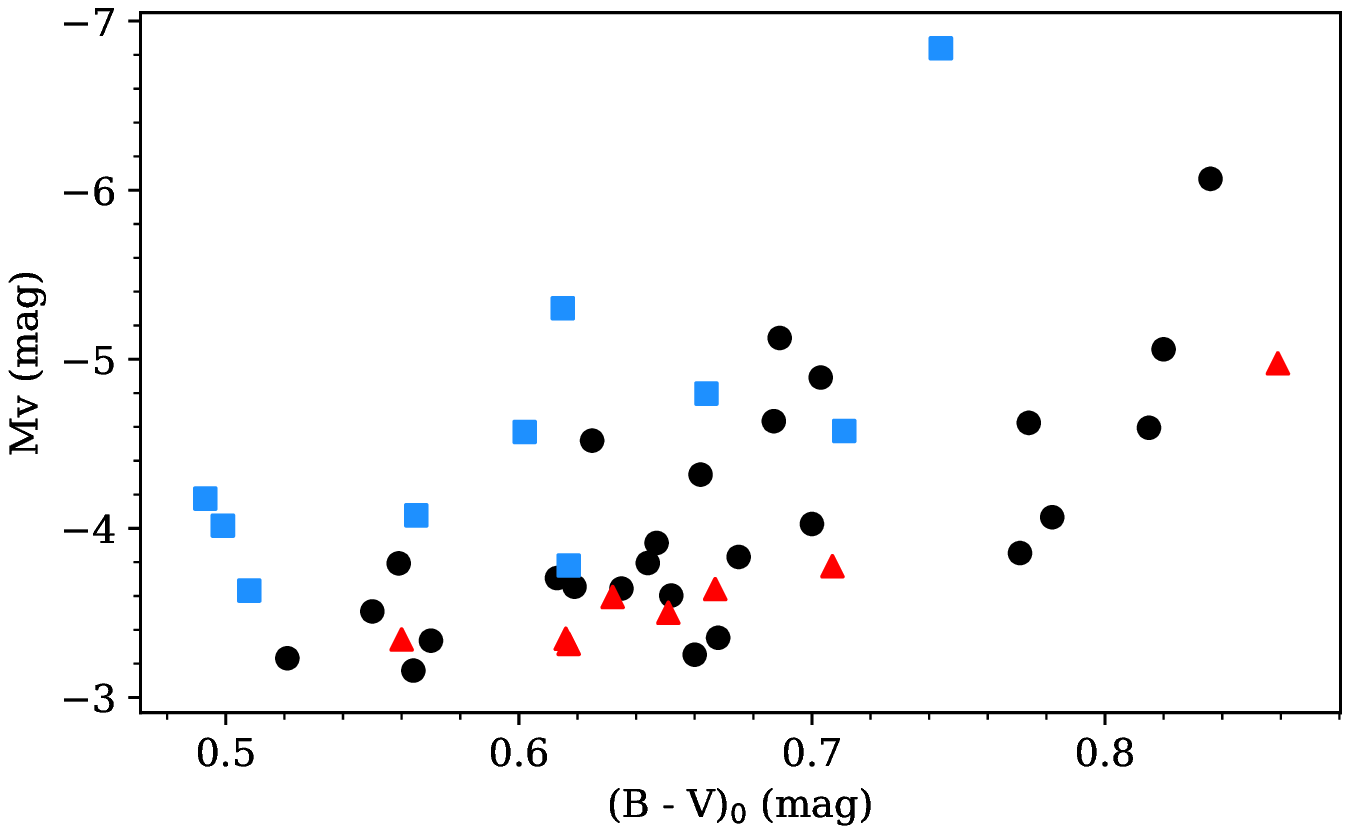}
\caption{\textbf{Left panel}: period-luminosity diagram in the V band. \textbf{Right panel}: color-magnitude diagram. Blue squares and red triangles indicate Cepheids outside one-sigma limit in the PL diagram. Black circles indicate Cepheids inside one-sigma limit. Sigma = 0.25 mag     \label{fig:fig8}}
\end{figure*}

\begin{deluxetable*}{ccc}
\tablenum{3}
\tablecaption{Period-radius relations in the form \\$log(R) = a_r \cdot  log(P)  + b_r$  \label{tab3}}
\tablewidth{0pt}
\tablehead{
\colhead{References} & \colhead{$a_r$} & \colhead{$b_r$}
}
\startdata
\cite{Sachkov} & 0.62  $\pm$ 0.03 & 1.23  $\pm$ 0.03 \\
\cite{Bono} & 0.655 $\pm$ 0.006 & 1.188 $\pm$ 0.008 \\
The present work [0.00; 1.00] & 0.66 $\pm$ 0.03 & 1.17 $\pm$ 0.03 \\
\cite{Petroni} & 0.676 $\pm$ 0.006 & 1.173 $\pm$ 0.008 \\
The present work [0.00; 0.85] & 0.68 $\pm$ 0.03 & 1.14 $\pm$ 0.03 \\
\cite{Gallenne} & 0.684 $\pm$ 0.007 & 1.135 $\pm$ 0.002 \\
\cite{Groenewegen+2007} & 0.686 $\pm$ 0.036 & 1.134 $\pm$ 0.034 \\
\cite{Turner+2002} & 0.747 $\pm$ 0.028 & 1.071 $\pm$ 0.025 \\
\cite{Molinaro} & 0.75 $\pm$ 0.03 & 1.10  $\pm$ 0.03 \\
\cite{Kervella2} & 0.767 $\pm$ 0.009 & 1.091  $\pm$ 0.011 \\
\cite{Storm+2004} & 0.77 $\pm$ 0.02 & 1.05  $\pm$ 0.03 \\
\enddata
\end{deluxetable*}

\begin{deluxetable*}{cccc}
\tablenum{4}
\tablecaption{Period-luminosity relations in the form $M_v = a_v \cdot  (log(P) - 1)  + b_v$  \label{tab4}}
\tablewidth{0pt}
\tablehead{
\colhead{References} & \colhead{$a_v$} & \colhead{$b_v$} & \colhead{Method}
}
\startdata
\cite{Groenewegen+2018}&-2.243 $\pm$ 0.137&-4.083 $\pm$ 0.118&Trigonometric parallax \\
\cite{Benedict}&-2.43 $\pm$ 0.12& -4.05 $\pm$ 0.02&Trigonometric parallax  \\
The present work [0.00; 1.00]&-2.50 $\pm$ 0.18 & -4.22 $\pm$ 0.05&ML \\
The present work [0.00; 0.85]&-2.67 $\pm$ 0.16 & -4.14 $\pm$ 0.05 &ML \\
\cite{Storm+2011}&-2.67 $\pm$ 0.10 & -3.96 $\pm$ 0.03&IRSB \\
\cite{Fouque}&-2.678 $\pm$ 0.076 & -3.953 $\pm$ 0.023&IRSB \\
\cite{Gieren}&-2.690 $\pm$ 0.100 & -3.981 $\pm$ 0.033&IRSB \\
\cite{Kervella2}&-2.769 $\pm$ 0.073 & -4.209 $\pm$ 0.075&IRSB \\
\cite{Molinaro}&-2.78 $\pm$ 0.11 & -4.20 $\pm$ 0.11&CORS \\
\cite{Turner+2010}&-2.78 $\pm$ 0.12 & -4.07 $\pm$ 0.10&Cepheids in clusters\\
\cite{Anderson}&-2.88 $\pm$ 0.18 & -3.90 $\pm$ 0.16&Cepheids in clusters\\
\enddata
\end{deluxetable*}

In the absence of observational and methodological uncertainties the scatter of PL diagram represents the finite width of the instability strip in color-magnitude diagram as the upper and lower envelope lines of the PL relation are the traces of the blue and red color boundaries of the instability region (\citealt{Sandage}). Left panel of Figure~\ref{fig:fig8} shows period-luminosity diagram in which blue squares (red triangles) indicate relatively bright (faint) Cepheids. Color-magnitude diagram with the same objects and designations is depicted on the right. As expected, most of the bright Cepheids are on the blue edge. On the other hand, the Cepheids located lower the calculated PL relation don't represent the red edge. We don't yet have a clear explanation of this behavior, it could be a result of some complicated evolutionary features as well as perturbations arising due to star-star interactions, since six out of eight faint Cepheids belong to binary/multiple systems.

\section{Discussion} \label{sec:discussion}
\subsection{The period-radius relation} \label{sec:pr}

\begin{figure*}[ht!]
\includegraphics[height= 10.0cm]{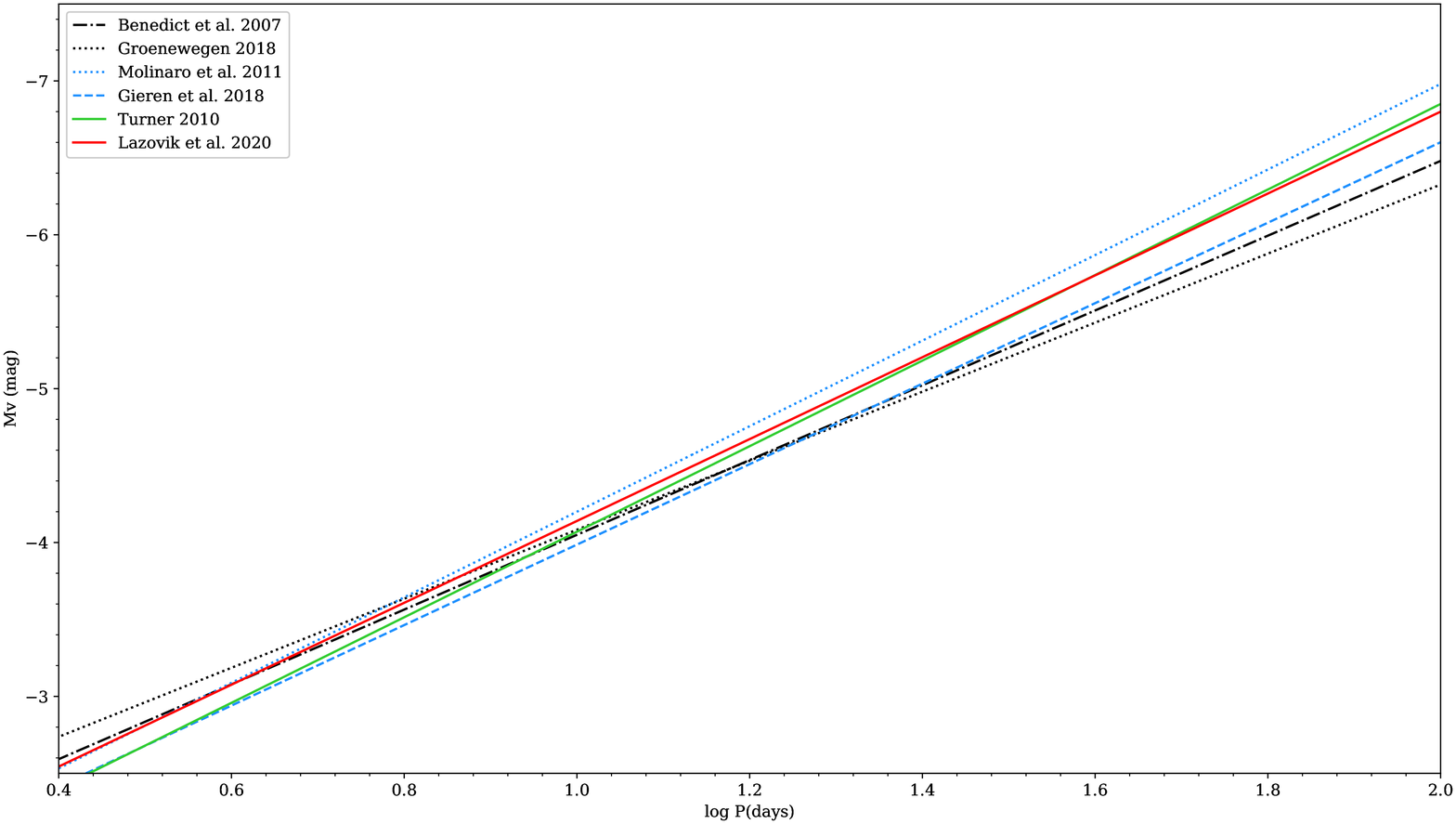}
\caption{Period-luminosity relations obtained by different authors. Black color represents relations based on trigonometric parallax method (black dashed-dotted line: HST parallaxes (\citealt{Benedict}); black dotted line: Gaia DR2 parallaxes (\citealt{Groenewegen+2018})). Blue color represents modifications of Baade--Becker--Wesselink method (blue dotted line: CORS method (\citealt{Molinaro}); blue dashed line: IRSB method (\citealt{Storm+2011})). Green solid line represents PL relation obtained using Galactic Cepheids in open clusters and groups by \cite{Turner+2010}. Red solid line represents PL relation from the present work (obtained with phase limitation).       \label{fig:fig9}}
\end{figure*}

Our PR relation is in good agreement with the results of the previous studies (see Table~\ref{tab3}). In particular, \cite{Gallenne} applied SPIPS algorithm, which is an implementation of the BBW technique, to 29 LMC and 10 SMC Cepheids in order to derive all their main parameters and to calibrate the projection factor and the PR relation. The obtained coefficients are consistent with our estimations. \cite{Groenewegen+2007} derived very similar PR relation after investigating five stars with known distances and angular diameters as a function of the pulsation phase. In contrast, some studies (\citealt{Turner+2002, Storm+2004,Kervella2, Molinaro}) propose steeper PR relation. In our opinion, the difference is related to the impact of the projection factor. We note that the first two works are based on the approaches that allow to independently estimate p-factor, while in the latter four works p-factor was adopted from other studies. Theoretical studies (\citealt{Bono, Petroni}) also confirm our results. At the same time, \cite{Sachkov}, applying Balona's method to the large sample of 62 Galactic Cepheids, computed a slightly shallower slope.

\subsection{The period-luminosity relation} \label{sec:pl}
Our results are compared with different studies in Table~\ref{tab4}. Moreover, for better visualization we plotted our PL relation with main relations from Table~\ref{tab4} in Figure~\ref{fig:fig9}. Each color represents the applied approach: black color corresponds to trigonometric parallax, blue color corresponds to the Baade--Becker--Wesselink method and green color corresponds to the expression based on Cepheids in open clusters, our PL relation is highlighted in red. The results achieved using different methods and techniques are not self-consistent, there is a considerable scatter in the PL diagram, with trigonometric relations being shallower than other relations. As we pointed out in Section~\ref{sec:intro}, the precision of trigonometric parallax decreases rapidly with distance. Both the zero point and the slope of the PL relations are correlated with the assumed parallax zero-point offset, whose influence doesn't allow to improve the existing quality of distance scale calibration (\citealt{Groenewegen+2018}). Nevertheless, in the range of small periods, which is related to predominantly close Cepheids with higher astrometric precision, our PL relation is in reasonable agreement with trigonometric parallaxes, especially with HST parallaxes (\citealt{Benedict}) Applying the ML technique generally leads to the brighter Cepheids and therefore to the longer distance scale, as compared with the results provided by the IRSB method (\citealt{Kervella2, Fouque, Storm+2011, Gieren}) The latter is explained by the differences between the ML and IRSB techniques highlighted in Section~\ref{sec:method}. The best match is found between our relation and the relation from \cite{Molinaro}.

The obtained results have global cosmological meaning, as the longer distance scale indicates the lower value of the Hubble constant, $H_0$. The current value of the Milky Way calibration used by \cite{Riess+2019} relies on HST parallaxes (\citealt{Benedict}). The zero point of the corresponding PL relation differs by 0.09 mag from the results presented in this study. Adopting such offset in the Cepheid calibration would reduce the Hubble constant from $H_0 = 74$ km/s/Mpc to $H_0 = 71$ km/s/Mpc. However, we would like to caution the reader against taking this estimate too seriously, as it is based on a simplifying assumption that all Cepheids are 0.09 mag brighter than was previously thought. Our main goal here is to demonstrate that there is evidence that supports the idea that the local value of the Hubble constant has to be lower. Future investigations are needed to support or disprove it.

\subsection{The distances} \label{sec:dist}
\begin{figure}[ht!]
\plotone{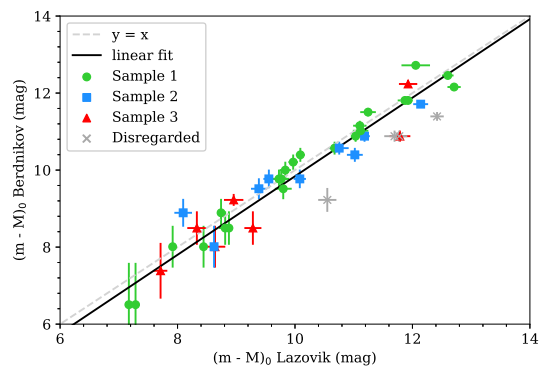}
\caption{A comparison of the obtained distances and the distances from \cite{Melnik}. Marker designations are the same as in Figure~\ref{fig:fig7}, grey crosses represent the overtone Cepheids which have been disregarded in the fit for reasons mentioned in the text. Grey dashed line: y = x. Black solid line: linear fit.  \label{fig:fig10}}
\end{figure}

Our distance estimates are in excellent agreement with data from \cite{Melnik}. Their procedure of calculating distances is based on the K-band period-luminosity relation from \cite{BerdnikovVozyakova1} and interstellar-extinction law derived by \cite{BerdnikovVozyakova2}. The distances from both studies, plotted against each other in Figure~\ref{fig:fig10}, are almost identical for the majority of Cepheids from the first two samples and even for several Cepheids from the third sample. Four suspected overtone Cepheids, namely Y Lac, DL Cas, RS Ori, SU Cyg, have been disregarded in the linear regression because in \cite{Melnik} they are identified as the fundamental pulsators. The convergence is expressed by the following linear fit:\\ $(m - M)_{Berdnikov} = (1.02 \pm 0.05) \cdot (m - M)_{Lazovik} - (0.36 \pm 0.43)$,\\ which is indistinguishable from $y = x$ relation within the margin of error. The perfect match confirms the accuracy of our color excess evaluations. It is worth noting that the K-band extinction is negligible, while V-band extinction may be significant.

\subsection{The reddening system} \label{sec:CE}

As mentioned in Section~\ref{sec:method}, independent and direct reddening determinations for individual Cepheids are one of the main advantages of the ML technique over the other modifications of the BBW method. Nowadays the most conventional way to derive color excess implies observations of the early-type reference stars. These objects are plotted on the color-color diagram and compared with the relationships between spectral type and intrinsic color for standard (non-rotating, ZAMS) stars to obtain the color excess, which is subsequently transformed to fit the Cepheid spectral type.  

\begin{figure}[ht!]
\plotone{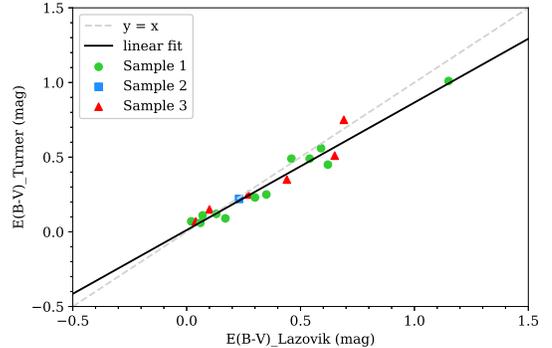}
\caption{A comparison of the obtained color excesses with reddening compilation by \cite{Turner+2016}. Marker designations are the same as in Figure~\ref{fig:fig7}. Grey dashed line: y = x. Black solid line: linear fit.  \label{fig:fig11}}
\end{figure}

The reddening system of \cite{Turner+2016} was established by the algorithm described above, using the \cite{Fernie} transformation. A comparison of the reddenings derived in the present study with color excesses by \cite{Turner+2016} is depicted in Figure~\ref{fig:fig11}. There is moderate scatter as well as reddening-dependent trend, which is indicated by the following regression fit:

$(B-V)_{Turner} = (0.85 \pm 0.02) \cdot (B-V)_{Lazovik} + (0.05 \pm 0.02)$

We can't be confident about where these discrepancies originate, but in our opinion the results of the ML technique are more reliable since our approach is more straightforward and based on the minimal amount of the initial assumptions. There is no critiquing of the study by \cite{Turner+2016}, however, we would like to point out the possible reasons for the inconsistencies in the color excess estimates.

First of all, we think that the \cite{Fernie} reddening transformation should be updated. The calibration of a quantity that defines reddening of a late-type star relative to that of an early-type star is based on a scarce number of data points and doesn't take into account a possible dependence on metallicity. Secondly, the assumption that the reference stars can be approximated by zero-age zero-rotation main-sequence standard stars at solar metallicity is likely to be invalid in some cases. It is important to remember that B0-stars have relatively short main-sequence lifetimes, that's why a fraction of the reference stars may be evolved stars with different color-color relation. The effect of rotation is also significant, especially for intermediate- and high-mass stars, as it impacts the efficiency of transport and mixing of chemical elements, modifying the internal stellar structure (\citealt{Heger1,Heger2,Mathis,Deal}). Finally, the determination of the reddening law appears to be an essential and complicated task. The study by \cite{Cardelli} indicates large systematic differences in extinction for lines of sight with considerably different values of $R_v$, spanning from 2.60 to 5.60. The enormous range of properties exhibited by UV extinction in the Milky Way is also discussed by \cite{Fitzpatrick1} and \cite{Fitzpatrick2}. It has been found that the properties of Milky Way extinction are not well-determined (\citealt{Fitzpatrick1}) and  IR-through-UV Galactic extinction curves have too large scatter to be considered as a simple 1-parameter family (\citealt{Fitzpatrick3}). Moreover, even for a given cluster the range of $R_v$ values can be broad. 

It would be informative to dwell on two objects for which the biggest deviations in color excess are found, namely S Vul and SV Vul (the corresponding deviations are 0.14 and 0.17 mag, respectively). These objects are the long-period Cepheids which were previously thought to be located in two star-forming associations, namely Vul OB1 and Vul OB2. Such associations, as well as the embedded clusters and groups, are commonly characterized by the substantial differential extinction, which might cause errors in the derived space reddenings. Recent studies (\citealt{Negueruela}) suggest that SV Vul is a member of another cluster, namely Alicante 13. Their results cast doubt on the classical view of two separate associations, Vul OB1 and Vul OB2, projected over the same region. Instead, supposed members of Vul OB1 and Vul OB2 may be distributed over the wide range of distances, making the classical approach of reddening estimation even more complicated. 

\begin{figure*}[htp]
\gridline{\fig{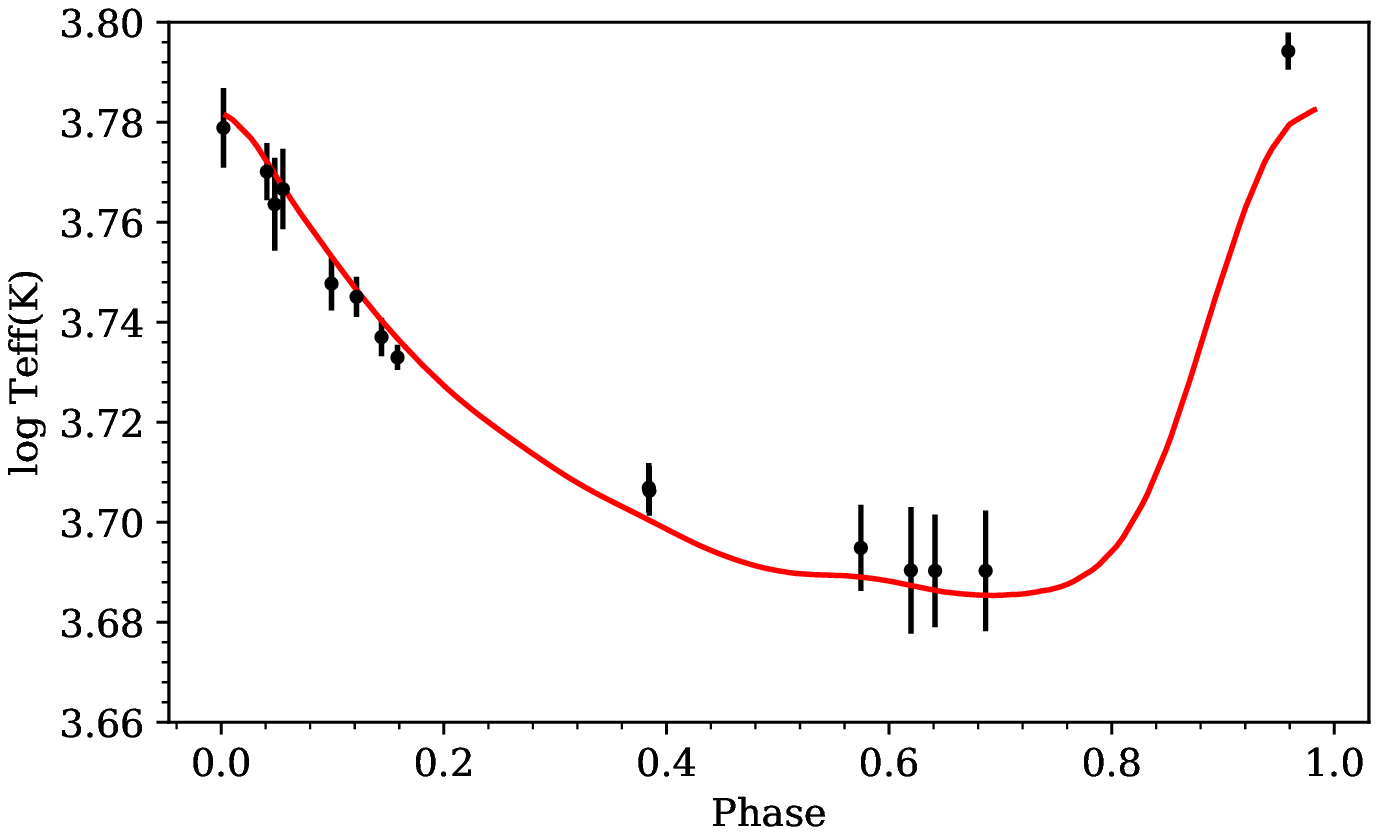}{0.45\textwidth}{}
          \fig{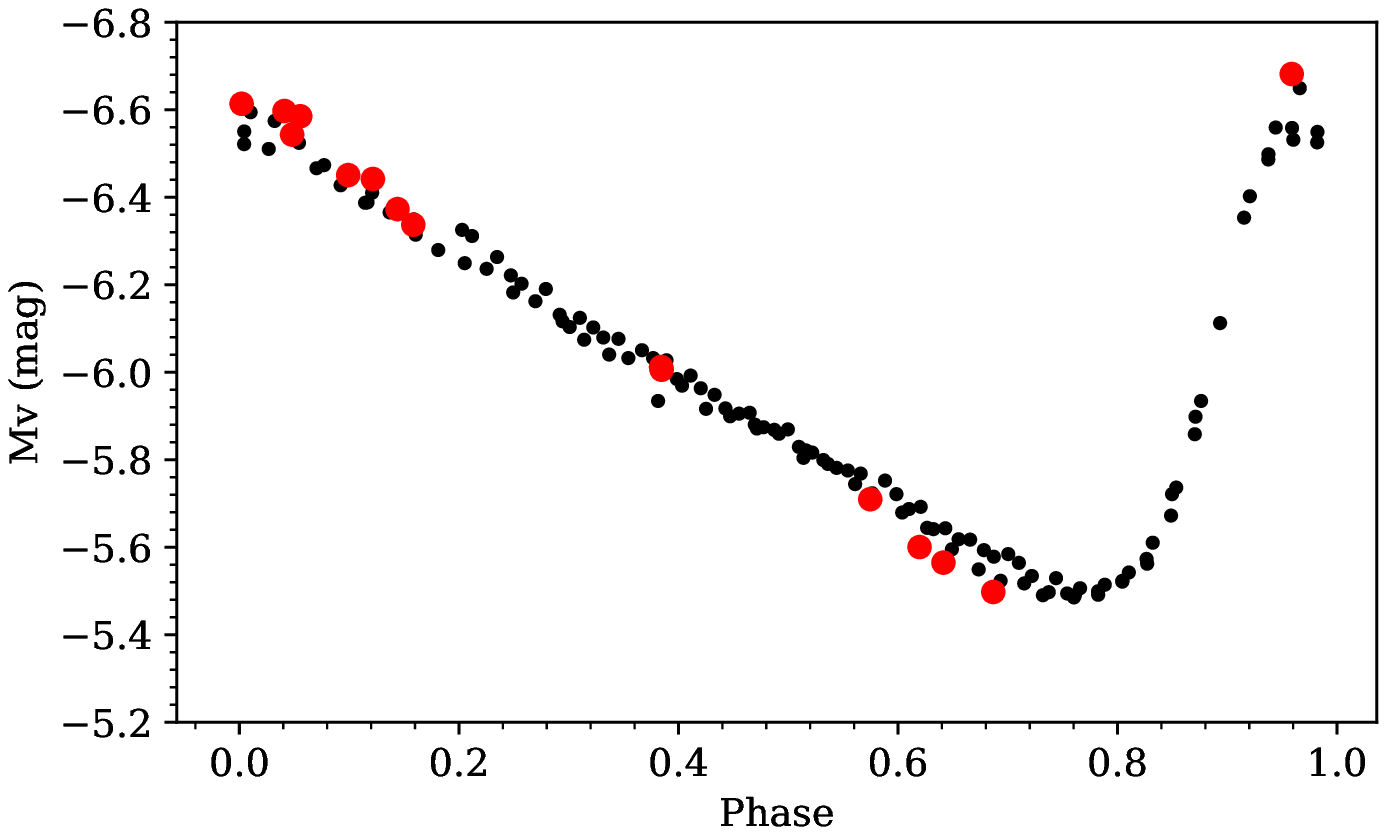}{0.45\textwidth}{}}
\gridline{\fig{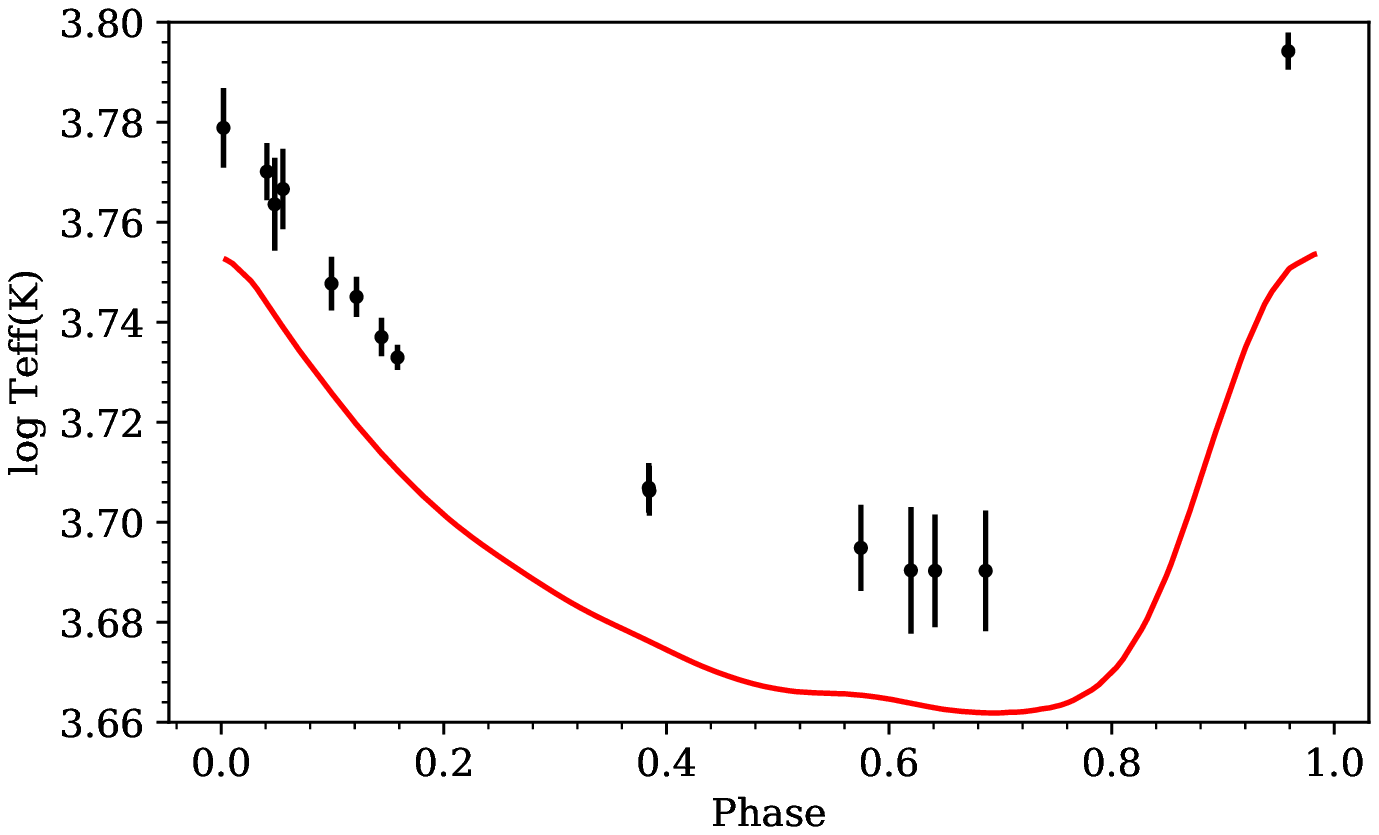}{0.45\textwidth}{}
          \fig{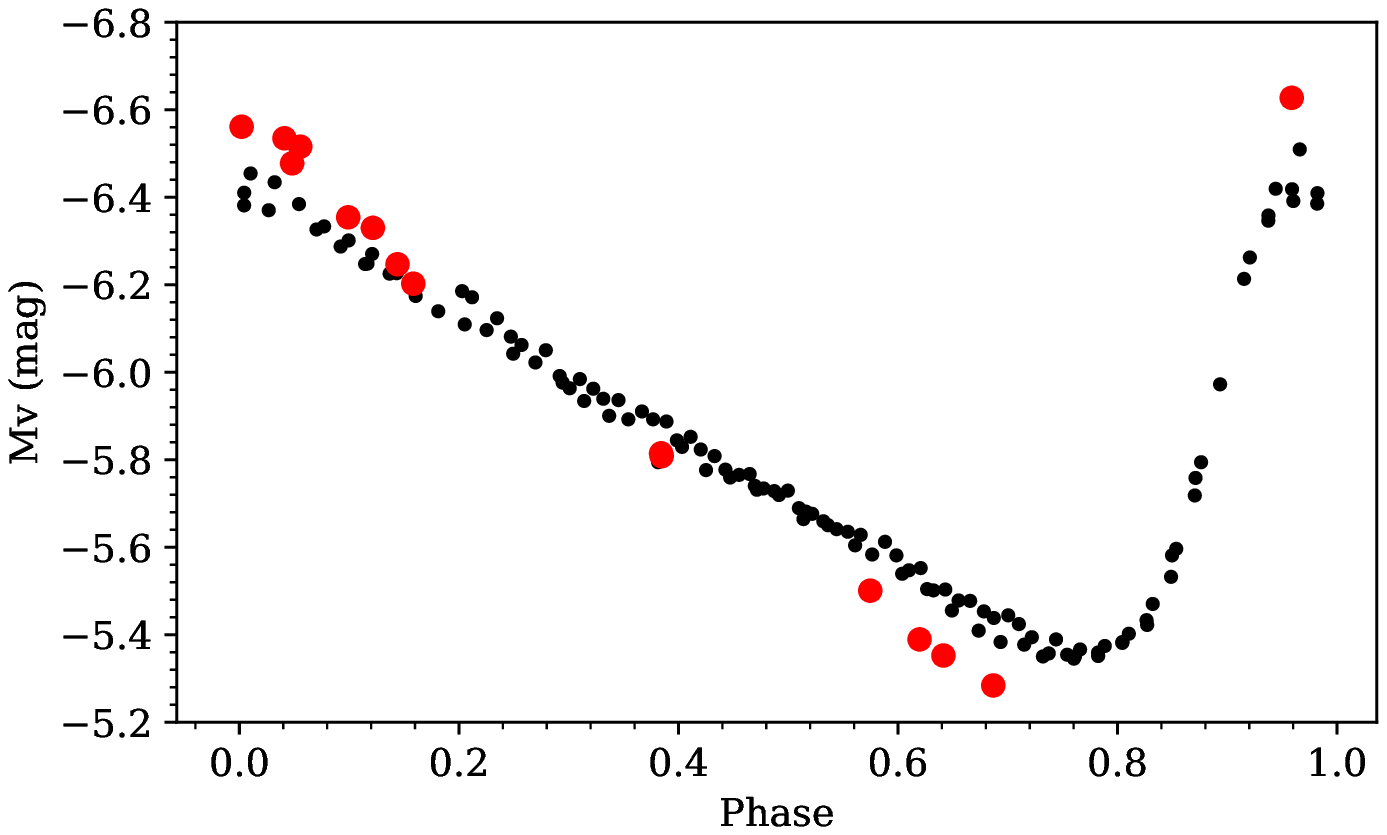}{0.45\textwidth}{}}

\caption{\textbf{Left column}: Effective temperature curves for SV Vul. Marker designations are the same as in Figure~\ref{fig:fig1}.  \textbf{Right column}: light curves for SV Vul. Marker designations are the same as in Figure~\ref{fig:fig2}. \textbf{Top row} corresponds to $E(B-V) = 0.62$ from the present study. \textbf{Bottom row} corresponds to $E(B-V) = 0.45$ from \cite{Turner+2016}. \label{fig:fig12}}
\end{figure*}

The sensitivity of the effective temperature curve to the color excess variation in the case of SV Vul is illustrated on the left hand side of Figure~\ref{fig:fig12}. Adopting the reddening value from \cite{Turner+2016} leads to the discrepancy between the calibrated and observed values of the effective temperature which is larger than the margin of errors. Moreover, the shape of the light curve calculated  from Eq.(\ref{eq1}) with the color excess from \cite{Turner+2016} doesn't reproduce the shape of the observed light curve as good as the analogous curve calculated with the color excess from the present work (the right hand side of Figure~\ref{fig:fig12}). The choice of reddening doesn't affect radius estimation, since the Balona equation (Eq.~\ref{eq8}) includes only the observed values of color index. Adopting $E(B-V) = 0.45$ from \cite{Turner+2016} will slightly change the value of absolute magnitude from $\overline{M_v} = -6.08 \pm 0.05$ to $\overline{M_v} = -5.94 \pm 0.10$.

At the same time, S Vul and SV Vul, like the majority of long-period Cepheids, undergo fast period variations, that's why careful data reduction has to be provided to obtain smooth phase curves. Several data sets have been removed to make the remaining data converge with the single period value. For this reason, period variation arises as an extra source of errors in the ML determination of reddening, whose impact is difficult to assess.

\section{Summary} \label{sec:summary}
In this study we have demonstrated the main features of the maximum-likelihood technique, which was originally developed by \cite{Balona+1977} and then modified by \cite{Rastorguev+Dambis+2010}. The method combines effective temperature data with light, color, and radial velocity variations to determine the amount of interstellar reddening, to compute the key parameters of Cepheids, including radius and absolute magnitudes, and to estimate the absolute distance modulus. Applying this method to 44 Galactic Cepheids allows us to obtain the following period-radius and period-luminosity relations: 

$log\,R = (0.68 \pm 0.03) \cdot log\,P + (1.14 \pm 0.03)$,

$M_v = - (2.67 \pm 0.16) \cdot (log\,P - 1) - (4.14 \pm 0.05)$.

Our results are generally in good agreement with previous works. The period-radius relation is confirmed by theoretical studies (\citealt{Petroni}) and consistent with empirical works (\citealt{Gallenne, Groenewegen+2007}). The period-luminosity relation supports findings from \cite{Molinaro} and is compatible with HST parallaxes (\citealt{Benedict}), although it supports a slightly brighter Cepheid calibration and thus a larger distance scale than the IRSB technique and HST parallaxes.

The possibilities of the ML technique are far from being exhausted. Our next steps depend on future data. As for today, the biggest limitation arises from the number of Cepheids with multiphase effective temperature measurements but, in the future, we will expand this list for higher precision of the obtained relations. This is relevant to long-period variables because at the moment the group of Cepheids in our research contains only five objects with the fundamental period exceeding 20 days. The lack of such objects prevents us from reducing the uncertainty in the final PR and PL relations. We would like to apply our approach to SMC and LMC Cepheids to show that our method can reach extragalactic objects. Using multi-band photometry will open new prospects for moving toward and improving the quality of distance scale calibration.

\acknowledgments

We are grateful to the Russian Foundation for Basic Research for
partial financial support (projects No. 18-02-00890 and
19-02-00611).

\bibliography{MLCepheids}{}
\bibliographystyle{aasjournal}
\end{document}